# An Alternative Approach to the Calculation and Analysis of Connectivity in the World City Network

S. Hennemann[a] and B. Derudder[b]

[a] Justus-Liebig-University of Giessen, Dept. of Geography, Senckenbergstrasse 1, D-35390 Giessen, Germany
[b] Ghent University, Dept. of Geography, Krijgslaan 281/S8, B9000 Gent, Belgium

**Abstract**
Empirical research on world cities often draws on Taylor's (2001) notion of an 'interlocking network model', in which office networks of globalized service firms are assumed to shape the spatialities of urban networks. In spite of its many merits, this approach is limited because the resultant adjacency matrices are not really fit for network-analytic calculations. We therefore propose a fresh analytical approach using a primary linkage algorithm that produces a one-mode directed graph based on Taylor's two-mode city/firm network data. The procedure has the advantage of creating less dense networks when compared to the interlocking network model, while nonetheless retaining the network structure apparent in the initial dataset. We randomize the empirical network with a bootstrapping simulation approach, and compare the simulated parameters of this null-model with our empirical network parameter (i.e. betweenness centrality). We find that our approach produces results that are comparable to those of the standard interlocking network model. However, because our approach is based on an actual graph representation and network analysis, we are able to assess cities' position in the network at large. For instance, we find that cities such as Tokyo, Sydney, Melbourne, Almaty and Karachi hold more strategic and valuable positions than suggested in the interlocking networks as they play a bridging role in connecting cities across regions. In general, we argue that our graph representation allows for further and deeper analysis of the original data, further extending world city network research into a theory-based empirical research approach.

## 1 Introduction

Peter Taylor's (2001) formal specification of the 'world city network' (WCN) as an 'interlocking network' has been a milestone in quantitative research on the geographies of globalized urbanization. In this 'interlocking network', cities are deemed to be connected through the flows of information, knowledge, capital, etc. generated within the office networks of advanced producer services (APS) firms. The popularity of Taylor's 'interlocking world city network model' (IWCNM) can be traced back to the combination of (i) its firm theoretical grounding (i.e. he uses Sassen's (1991) well-known work on 'the global city' as a starting point, and draws on a specification that is well established in social network analysis), and (ii) the fact that the empirical world/global city literature before the turn of the century was flawed by a combination of eclecticism and fuzziness. Drawing on this IWCNM specification and a series of concomitant data gatherings to 'feed' the model, Taylor and his colleagues from the Globalization and World Cities (GaWC) research network have provided detailed descriptions of the geographical contours of the WCN (e.g. Taylor et al, 2002; 2010)[1]. Collectively, these accounts have most certainly enhanced our understanding of the outline of the WCN.

---

[1] A number of other 'agents' have been analysed based on this methodological framework (e.g. Taylor, 2005; Hoyler and Watson, 2012), but Taylor's initial specification as well as most of GaWC's analyses are based on APS firms.

In recent years, however, urban network research drawing on the IWCNM has come under scrutiny. We are hereby not referring to a series of postmodern critiques that take issue with the model because it purportedly represents a totalizing metanarrative (see Robinson, 2002; Smith, 2012), but rather to researchers who have positively engaged with the IWCNM in order to extend its remit. The purpose of this paper is to contribute to this literature by devising a method that allows obtaining insight into the significance of relations and positions of cities in the office networks of APS firms, rather than merely taking IWCNM-produced results at face value. To this end, we introduce a measurement framework based on the betweenness centrality of cities in a revamped specification of the WCN, whereby the measures are interpreted against a randomized baseline model that retains the network's original degree distribution. The advantages of this framework are twofold: (1) betweenness centrality is a more useful gauge of network positionality than mere IWCNM connectivity as the latter only measures 'local' network effects; while (2) using a baseline model allows distinguishing between 'meaningful' connectivities and linkages on the one hand and those that can merely be attributed to random chance on the other hand.

The remainder of this paper is organized as follows. We first review the IWCNM and the way in which researchers have used its shortcomings to call for extending/altering the model beyond its initial specification in Taylor (2001). We use this overview to detail the construction of a different projection function for obtaining inter-city networks based on information on the office networks of APS firms, after which we introduce our analytical framework, which consists of randomized baseline model for gauging betweenness centrality in networks. In the following section, we discuss our results, and explain how these demonstrate how our approach may enrich quantitative WCN research. The paper is concluded with an overview of our main findings and some potential avenues for further research.

## 2 Beyond the IWCN model

### 2.1 The IWCN model

Before Taylor's (2001) specification of the IWCNM, empirical research on the shape of transnational urban networks simply relied on a variety of commonsensical indicators, such as cities' roles as the headquarter location for multinational enterprises and international institutions (e.g., Godfrey and Zhou, 1999), their insertion in global transport networks (e.g., Keeling, 1995), etc. Although most of the resulting 'world city rankings' were inherently plausible, as a genuine empirical framework they all shared an obvious deficit, i.e. the lack of a *precise specification of what constitutes the key dynamics behind WCN-formation*. However, as Taylor (2001, p. 181) put it, such a precise specification is imperative when studying the WCN, because "(w)ithout it there can be no detailed study of its operation - its nodes, their connections and how they constitute an integrated whole." Drawing on Sassen's (1991) work on the rise of globalized advanced producer services (APS) economies in New York, London and Tokyo, Taylor (2001) thus specified the WCN as an inter-locking network with three levels: a network level (the global economy), a nodal level (world cities), and a critical sub-nodal level (firms providing the APS). According to Taylor's specification, it is at the latter level that WCN formation takes place: through their attempts to provide a seamless service to their clients across the world, financial and business service firms have created global networks of offices in cities around the world. Each office network represents a firm's urban strategy for servicing global capital, and the WCN can therefore be formally quantified by analysing the aggregated geographical patterns emerging from the flows within the office networks of such firms.

Taylor's (2001) formal specification of the WCN as an interlocking network starts with a universe of $m$ advanced producer service firms in $n$ world cities. The importance of the office of a firm $j$ in city $i$ is measured through its 'service value' $v_{ij}$, which can be arrayed as a service value matrix $V_{ij}$. The basic connection $r_{ab,j}$ between each pair of cities $a$ and $b$ in terms of a firm $j$ is derived from the initial matrix $V_{ij}$ as follows:

$$r_{ab,j} = v_{aj}.v_{bj} \qquad (1)$$

The conjecture behind conceiving the product of the service values – basically an extremely simple interaction model – as a surrogate for actual flows of inter-firm information and knowledge between cities is that co-location of firms in two cities opens up the possibility of an actual link between both cities[2]. Based on this elemental link, a host of related measures can be devised. Perhaps the most commonly used measure in GaWC research is a city's 'global network connectivity' *GNC*, which is obtained by aggregating all basic relational elements across all cities and across all firms:

$$GNC_a = \sum_{b,j} r_{ab,j} \qquad (2)$$

As can be gleaned from the above, the 'measurement' of the WCN based on the IWCNM requires information on the presence of a large number of globalized APS firms in a large number of cities (Taylor et al, 2012). For instance, in GaWC's most recent data gathering, which will also be used as the input to our calculations, the (importance of the) presence of 175 globalized APS firms in 525 cities across the globe is gauged through 175 x 525 = 91875 service values $v_{ij}$. The latter data is garnered by focusing on two features of a firm's office(s) in a city as shown on their corporate websites: first, the size of office (e.g. number of practitioners), and second, their extra-locational functions (e.g. regional headquarters). Information for every firm is thereby simplified into service values ranging from 0 to 5 as follows. The city housing a firm's headquarters is scored 5, while a city with no office of that firm is scored 0. An 'ordinary' or 'typical' office of the firm results in a city scoring 2. With something missing (e.g., no partners in a law office), the score is reduced to 1. Particularly large offices or national headquarters are scored 3, and those with important extra-territorial functions (e.g., regional headquarters) score 4 (Figure 1).

Using this service value matrix $V_{ij}$ as the input to equation (1), and aggregating this across all firms leads to a city-to-city adjacency matrix, which can be described by measures such as cities' *GNC* (equation 2), while the overall network $R_{ab}$ can be analysed by means of a host of multivariate techniques (e.g., principal components analysis in Taylor et al, 2002; fuzzy cluster analysis in Derudder et al, 2003) and/or network analysis techniques (e.g., clique analysis in Derudder and Taylor, 2005).

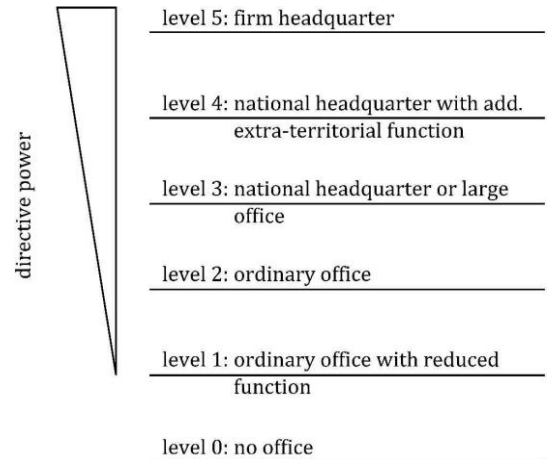

**Figure 1**: Hierarchization in the GaWC service value matrix.

**2.2 Problems with the IWCNM**

Although the IWCNM has been widely applied, it has not been without its critics. Here we describe two critical interventions that are crucial to appreciate our approach towards extending its empirical remit[3].

The first critique of the IWCNM has been that Taylor's specification summarized in equation (1) is tautological. Comparing a simple aggregation of service values with GNC measures derived from the IWNCM shows that there is often very little difference between both rankings, which has led Nordlund (2004) to reproach Taylor for "turning apples into oranges." Here we follow Taylor's (2004) reply to this critique, in which he argues that the alleged empirical parallels do not simply mean that the model has no added value in urban network terms. For instance, it is clear that in the IWCNM a city hosting a *large* number of APS firms with a *limited* global reach will not be well-connected, while a city hosting a *small* number of service firms with an *extensive* global reach will be well-connected. The fact that this is not always very visible empirically may simply be the by-product of some sort of autocorrelational sorting in the location strategies of firms (major firms choosing the same set of major cities), and is as such no reason to dismiss the IWCNM in and by itself. However, although in our view misguided, Nordlund's (2004) critique does raise the relevant question of the added value of the 'network measures' emanating from the IWCNM: very few GaWC analyses manage to add something beyond the obvious point that New York and London have large GNCs because both cities are housing a lot of globalized service firms. Put differently: actual network measures assessing cities' *position* in the overall network have not been widely adopted in GaWC research, and it can therefore be argued that their analyses have not always fully lived up to the inherent potential of Taylor's network specification.

---

[2] Although results depend on the designation and scope of the service values, experiments by Liu and Taylor (2011) have shown that this has little impact on the results per se. As a consequence, even if $v_{ij}$ is simply measured as a binary variable (0 = no presence, 1 = presence, so that there is only the possibility of an interaction if the service values for both cities are = 1), then the results are still in line with more subtle hierarchical operationalizations of $v_{ij}$.

[3] As suggested in the intro, we thereby only deal with those writings that assume that there 'is' a WCN that can be 'measured': given the cultural turn in (especially British) human geography, this has become a less obvious vantage point in WCN research, and the majority of the Taylor critics thus take issue with his desire to quantify more than anything else.

Second, it has recently been argued that the IWCNM is essentially a 'dumbed down' version of the much richer service value matrix. Taylor's (2001) specification of the WCN essentially starts off as a so-called 'two-mode network'. A two-mode network consists of two disjoint sets of nodes, whereby the primary data connects nodes of both sets. The WCN Taylor-style clearly begins as a two-mode network: it consists of two disjoint sets of nodes (cities and firms), whereby the primary data consists of links connecting nodes of the different sets (the presence of firms in cities). In principle, two one-mode networks can be projected from a two-mode dataset, and the IWCNM is essentially one of the possible ways of deriving a one-mode city-to-city network from the two-mode city-to-firm dataset (Liu and Derudder, 2012). In the network literature, it is agreed that two of the major problems associated with IWCNM-like projections of two-mode networks into one-mode networks include (1) information loss due to compression of the two-mode network, and (2) an inflation of linkages due to the inclusion of every possible pairwise link (Latapy et al, 2008). For our purposes, the first issue is not that relevant, as it basically draws attention to the fact that Taylor's focus on cities does not do full justice to the 'duality of cities and firms' (and which can be tackled by using two-mode network analytics, see Liu et al, 2012; Neal, 2008). Rather it is the second issue that is of concern here: the fact that equation (1) implies that the largest firm in the service value matrix enforces a wide-ranging density and therefore clustering on the IWCNM-derived network (Neal, 2012a; 2012b), making it very difficult to distinguish between methodological artefact and 'actual' results. Which relations are 'meaningful' and which relations are simply the by-product of the over-specification emanating from Taylor's one-mode network projection remains unclear. Importantly, the reason why GaWC researchers have largely drawn on multivariate techniques rather than engaging in an actual network analysis is related to this over-specification, as the IWCNM-produced networks are indeed very hard to handle empirically.

The net consequence of both limitations can clearly be seen in an analysis by Derudder and Taylor (2005), in which the authors attempt to move beyond GNC rankings derived from the IWCNM. That is, in their paper, Derudder and Taylor (2005) attempt to analyse the city-to-city network with 'genuine' network analysis techniques rather than standard multivariate techniques. In practice, the authors try to reveal the spatialities of the WCN by looking for 'cliques', i.e. groups of cities that closely interact with each other. Although clique analysis is a standard network analysis technique, the Derudder and Taylor (2005) paper shows above all that applying it in the context of the IWCNM proves to be very difficult. For instance, the very dense nature of the network specified by the IWCNM leads them to dichotomizing the data by imposing a number of thresholds. However, it is clear that such an approach leaves a lot to be desired for a number of reasons, not in the least because imposing an arbitrary threshold influences the results and does not do justice to the richness of the data. The key point for the present paper, however, is not so much the specific problems surfacing in this particular analysis. Rather, the point is that such difficulties are bound to re-emerge as long as the IWCNM connectivities are not re-specified in a way that makes them fit for network analysis (e.g. it is also impossible/meaningless to calculate betweenness centrality in a network where virtually every node is connected to every other node).

In this paper, we propose to tackle this issue as follows[4]: (1) we suggest an alternative one-mode projection of the two-mode network epitomized by the service value matrix $V_{ij}$, all the while retaining as much of the original relational information as possible. Compared to the initial specification, this involves wiping out as much as possible 'noise' in the data, which then opens up the possibility for (2) an actual network analysis of the WCN produced through the aggregated location strategies of APS firms. Here we will focus on betweenness centrality for nodes that allows for assessing the information flows that pass through a city node. In line with the original objective of Derudder and Taylor (2005), the result is a *network* analysis in that these measures evaluate the influence of a given node over the information flows in a network (cf. Gilsing et al, 2008). The next sections consecutively focus on this alternative specification of the WCN based on information contained in the service value matrix and a discussion of our network-analytical framework.

**3 Towards an alternative one-mode projection**

In this section, we discuss our specification of an alternative one-mode city-to-city adjacency matrix $R_{ab}$ based on the two-mode service value matrix $V_{ij}$. The aim is to arrive at a reliable graph representation that retains most of the initial information in the service value matrix, while avoiding the typical high density of the IWCNM network projection[5]. Our alternative has three major features: (1) the number of linkages will be significantly reduced, thereby (2) imposing directionality in the connections ($r_{ab} \neq r_{ba}$), and (3) assuming spatiality in the organization of APS firms' office networks.

In the IWCNM epitomized by equation (1), information on the relative importance of offices is used to guesstimate the strength of city-to-city connections for that firm. A major consequence is that each city-pair that shares offices of a certain firm is deemed connected, whereby there is furthermore no distance decay effect (Euclidean, functional, or otherwise) in the importance of the edges. For instance, if a firm has a 'national headquarter' in Amsterdam (service value 3) and 'typical offices' in Rotterdam and Karachi (service value 2), then these three cities are not only deemed to be inter-connected, but it will also be assumed that the Amsterdam-Karachi and Amsterdam-Rotterdam edges are of the same strength.

Our starting point to an alternative one-mode specification is that this is unlikely because most 'global' companies organize their business geographically through a territorial framework of sorts, such as countries and/or 'world regions' (see Figure 2 for the McKinsey example). This regionality in the organization of office networks is of course idiosyncratic for each firm, but here we assume a nested organization that consists of (a) countries, which are then further grouped in (b) a number of world regions that commonly return in transnational organizational schemes: North America, Central and South America, Europe, Africa/Middle East, West/Southwest Asia, and Pacific Asia/Oceania (see Figure 3)[6].

Here we combine this geographical classification with information on the hierarchization of office networks contained in the GaWC data to build a general algorithm assessing the primary linkage of cities within each office network[7].

---

[4] Although not necessarily cast in this form, these problems have recently given way to other attempts to devise alternative approaches to the measurement of border-crossing urban relations. In the context of the IWCNM's application in the analysis of polycentricity in metropolitan regions, Meijers et al (2012) propose to use a measure of 'related variety' in urban economies to measure interaction between cities-as-nodes. Closer in line with the original IWCNM, Neal (2012b) develops a statistical test for identifying linkages that are strong enough to suggest that they do not arise from random chance. In this paper, we develop a methodology that has a similar objective to that of Neal (2012b), but implement this in the context of network-analytical framework rather than sticking to the basic measures emanating from Taylor's (2001) specification.

[5] All data preparation and network-related calculation have been carried out using python/networkX for python (http://networkx.lanl.gov/).

[6] We emphasize that this coarse-grain approach to the geographical division of business network organization is of course prone to the problems associated with sweeping generalizations, but in addition to probably being more in line with 'actual' flows, it also has the critical advantage of retaining the nodal connectivities engendered in the IWCNM-produced WCN (see below).

[7] As an aside, it is worth mentioning that most GaWC analyses based on the IWCNM-derived adjacency matrix reveal strong regional tendencies in the WCN, which implicitly corroborates a regionalized approach to the WCN.

Our alternative specification essentially entails a geographical operationalization of the metaphor of 'reporting cascades' between offices in different cities. In the process, we are only considering reporting that goes 'upstream' from lower levels (i.e. lower service values) to higher levels (i.e. higher service values). As a consequence, connections between cities are only possible between offices offering different levels of servicing, and in a way that they are directed from the lower level towards the higher level. Referring back to our example, then, we assume that there are no information flows between Karachi and Rotterdam, while the direction of the 'reporting' will be from Rotterdam and Karachi to Amsterdam.

However, to this that we add our geographical component in that we assume that there is only one reporting city, defined by countries/world regions. For each city that houses an office of a firm, the algorithm connects this city to the city with the highest-level office within the same country. In case there is more than one office of this particular level in the same country, the connection is made with the geographically nearest city (e.g. connecting San Francisco to Los Angeles rather than to New York in case that Los Angeles and New York are at the highest service level within the United States). If there is no higher-level office in the same country, the connection is made to the highest-level office within the own world region. And if there is no such higher-level within a city's own world region, the connection is made with the global headquarters (service value 5). Referring back to our example, then, it can be seen that the Karachi-Amsterdam link is deleted (or to be more precise: not created), while the Rotterdam-Amsterdam link is retained (unless there would have been a more important office in the Netherlands than the one in Amsterdam). Figure 4 shows an example of the actual linking process for the company network of *McKinsey&Company*. The regionally bound sub-systems can be traced back to the small office categories 1 and 2, which mainly report to their regional headquarters, except in the case of US offices, which are directly connected to the New York headquarter. The regional hubs are connecting the lower level hierarchies to the higher order hierarchies, which resembles the global location that is reported on McKinsey's corporate website.

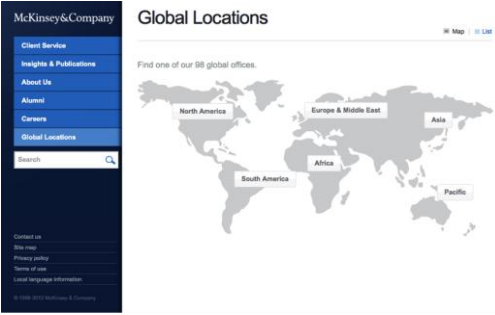

**Figure 2.** 'World regions' in McKinsey's office network.

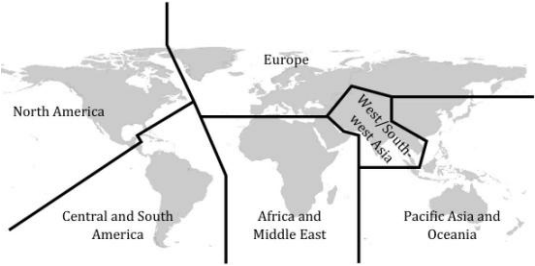

**Figure 3.** Assumed 'world regions' in typical APS firms' business organization.

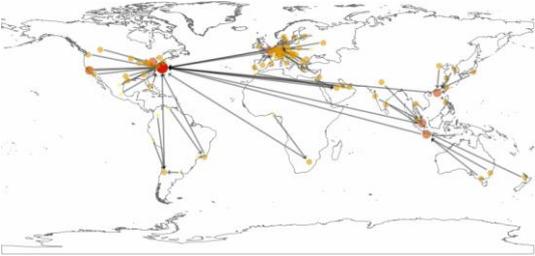

**Figure 4.** Outcome of the primary linkage procedure for McKinsey's firm network for illustrative purposes

Each of the APS firms' office networks is thus reduced to an empirically probable reporting graph, after which the final step basically involves aggregating the resulting networks across all firms so that edges become valued (e.g., if Amsterdam often houses the most important office of APS firms in the Netherlands, and Rotterdam also has a lot of offices of these firms, then there will be a sizable Rotterdam-Amsterdam link in the adjacency matrix).

This linkage procedure produces a one-mode city-to-city adjacency matrix that does not suffer from the over-specified nature of the IWCNM, while the end-result is a much sparser network that nonetheless retains the WCN's significant empirical structures. The latter appraisal is substantiated by an empirical comparison between a city's logarithm of the degree[8] based on our primary linkage algorithm and a city's GNC in the IWCNM: Figure 5 plots the relationship of both nodal connectivity representations[9]. The correlation coefficient of 0.95 indicates that both two models are similar with respect to what they (re)produce empirically in terms of basic structure and topology, and thus ensures that all further calculations – even those that go well beyond the IWCNM approach – reflect very similar systems. Put differently: we have produced a WCN that has similar properties to that produced by Taylor's IWCNM, but got rid of the over-specified nature that hampers its analyses through standard network-analytical tools. In addition, the graph represents a firm theoretical concept of command and control flows in global firm networks, therefore addressing the issues that have been raised by Nordlund (2004). The transformed adjacency matrix is now ready for an actual network analysis, which is described in the next section.

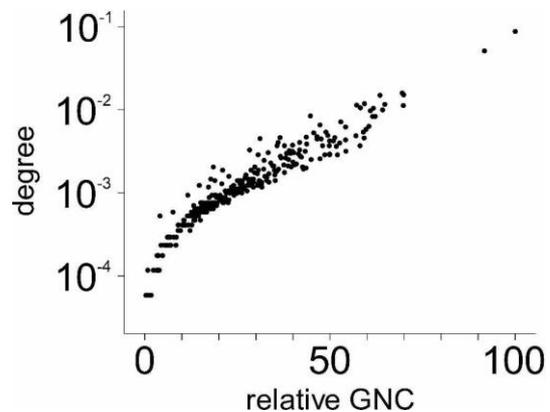

**Figure 5.** Relationship between the relative GNC and the logarithm of the degree of cities in the primary linkage graph.[10]

## 4 A randomization procedure for benchmarking betweenness centrality

Simple and largely descriptive approaches for evaluating complex systems such as WCNs are seldom offering insight into the significance of urban relations and positions. For instance, as stated in the introduction, although Taylor's WCN specification has helped the literature moving beyond empiricist guesstimates, it can be said that GaWC's GNC rankings merely reveal that a city is hosting a lot of important offices of a lot of important APS firms. Although this approach has some merits from a descriptive point of view, it lacks an actual appraisal of a city's overall position in the network set against a randomized reference network. To this end, we propose to compute betweenness centrality as a more refined measure of nodal prominence in the network, and use it to define prior expectations of the network structure by computing cities' (mean) baseline values.

---

[8] The degree centrality DC of a node v is defined as number of direct neighbors that connect to v. We also consider in-degree, which is the fraction of nodes that connect incoming edges to node v. Similarly, the out-degree is the fraction of nodes that connect outgoing edges to node v. The DC(v) and its directional variants are normalized by the total number of nodes N-1 in the network.

[9] Many network related measures and distributions follow power-law functions due to the complex nature of the underlying systems? In our case, this implies that very few cities are containing most of the activity (i.e. have a large connectivity), while most cities are not very active (i.e. have a small connectivity). The overall degree distribution of our network also follows a power law with an approximated exponent of -0.8 ($R^2$=0.88). This is a typical pattern for so-called scale-free networks, and reflects the nature of the underlying complex system with consequences for the topology and the organization of the network (cf. Barabási and Albert, 1999).

[10] Note: the degree in the directed and weighted network is the normalised aggregate of the in- and out-degree of the cities. The normalization accounts for the number of cities and firms in order to yield a value of 1.0 when all cities connect in and out to the given city for each of the individual firms.

Betweenness centrality is a network measure that has typically been lacking from GaWC analyses because of the over-specification resulting from the IWCNM. Overall, betweenness centrality gives a more refined appraisal of a node's centrality in a network as it assesses the number of shortest paths from *all* nodes to *all* other nodes passing through that node (if possible given the network structure). Betweenness centrality thus becomes a more useful measure of the position of a given node in the overall network than mere GNC: the latter can well be merely a local effect, while the former provides an assessment of node's importance to the entire network. The betweenness centrality $BC_a$ of a city *a* is given by the expression:

$$BC_a = \sum_{x \neq a \neq y} \frac{SP_{xy,a}}{SP_{xy}} \qquad (3)$$

whereby $SP_{xy}$ is the number of shortest paths from node *x* to node *y*, and $SP_{xy,a}$ is the number of these shortest paths running through *a*. However, as we are dealing with a weighted network, in practice links are considered in proportion to their capacity, which adds an extra dimension beyond the topological effects.

Defining a baseline assessment for betweenness centrality is complicated just as for any other network centrality measure, because stochastically independent generalized baseline models simply do not exist. For instance, the popular preferential attachment model of Barabási and Albert (1999), or similar models such as the small world model of Watts and Strogatz (1998), are both models that are often used as baseline models, because they capture some of the desired properties of a network and are defined only by very few parameters (e.g. the number of nodes and the average degree in the case of the BA-model). However, these models are considered to be too general in order to capture more subtle structures in real world networks that exhibit non-Gaussian broad distributions for most of their structural properties (cf. Andriani and McKelvey, 2009; van Wijk et al, 2010).

For our purposes, the main aim in devising a baseline model is to preserve the basic structural properties of the parameter distributions of the network, but to destroy (i.e. randomize) other properties. One of the most important single distributions for a network is the degree distribution, i.e. the frequency of the number of direct neighbours of nodes. A very efficient way to preserve this feature is to shuffle the edges by randomly picking two edges that connect four different nodes and swap their connections (Maslov and Sneppen, 2002). If this is repeated a sufficient number of times, the distribution of neighbours per node is still the same, but the original empirical relation between the nodes is effectively destroyed, and therefore independent from the empirical data. If this shuffling approach is combined with a bootstrapping, i.e. a repeated sampling of randomized networks, confidence intervals for the estimated parameter can be calculated (see Hennemann et al, 2012).

In our analysis, we implement an 'upper-level directed' randomization procedure, which preserves the empirical in-degree and out-degree distributions. Central to the swapping and at odds with the method proposed by Hennemann et al (2012) is that it is not completely destroying all relational dependency that is present in the empirical network. Instead, the edge direction is preserved, whereas the hierarchies and the geographical dependency are effectively lost after the randomization.

A concrete example may help the reader to appreciate the procedure and to evaluate the conceptual and empirical consequences of this approach. Consider two directed connections a-b and c-d, with connection a-b linking two low-level offices of an APS firm in South America (e.g. Bogota with $v_{ij} = 1$ and Sao Paulo with $v_{ij} = 2$) and connection c-d linking some high-level cities in Europe (e.g. Paris with $v_{ij} = 4$ and London with $v_{ij} = 5$). The swapping will produce two new connections, substituting the original ones in the random model, connecting a-d and c-b. In this example, the swap will connect Bogota with London, and Paris with Sao Paulo. The net consequences are that the geographical as well as the hierarchical dependencies are being 'destroyed', all the while retaining the overall out-degree/in-degree distribution. This swapping procedure is repeated ten times the total number of edges in the empirical network (approximately 32,000 swaps are conducted), which acts as starting point for the randomization and is successively shuffled with each iterative step of swapping. However, this randomized network represents only *one* possible random equivalent of the empirical network. In order to yield greater confidence on the difference between the empirical and the randomized network, we repeated this randomization for 100 configurations and estimate network measures from these random samples. This simulation or bootstrapping is discussed in Hennemann (2012) in detail. Comparing network centrality values derived from the empirical network with the mean betweenness centrality values in the randomized networks allows revealing *significant* patterns in the network positions of cities.

## 5. Results and discussion

Figure 6 plots the mean betweenness centrality in the randomized baseline networks against the empirical values derived from our linkage algorithm applied to the 2010 GaWC data. Note that all figures are double log plots: the relations are non-linear, but suggest the presence of power laws.

Comparing the first two plots (left and centre panel), it can be seen that our randomized model shows above all a fairly good fit to empirical in-degree. This implies that, given the network structure, cities 'receiving' reports have major centralities in the network at large, much more than in case of the out-degree. This can be interpreted as a post-hoc corroboration of the overall idea behind our alternative projection function procedure as well as being supportive for the reliability of the randomization process. Nonetheless, the mean values of the randomized betweenness and the empirical betweenness are only loosely connected (right panel; there is a minor power-law relationship at r = .47), which in turn suggests that benchmarking against the null/random model gives additional insight. The remainder of this section uses some of these differences between empirical betweenness, randomized betweenness, and IWCNM-produced GNC to explore how our empirical approach may enrich insight into the spatial structures of WCNs.

To explore some examples, Tables 1 and 2 compare cities' empirical betweenness to the GNC measures produced by the IWCNM. The ratio of the empirical betweenness connectivity and the mean of the randomized betweenness represents a normalized estimate for the betweenness, while significant deviations between observed and expected values were calculated on the basis of being two standard deviations from the mean (i.e. empirical betweenness scores are marked as significant when they deviate at least +/-2SD from the randomized means of the betweenness values in the randomized/simulated networks). Table 1 provides the results for the cities that feature in the top 20 of either empirical betweenness in our network specification or the GNC according to the IWCNM, Table 2 ranks the midfield of cities with average connectivity. In addition, Figure 7 plots the WCN at large, while Figure 8 plot the ego networks of a number of notable cases (i.e. cities with a much higher/lower empirical betweenness than GNC, such as Tokyo, Melbourne, Karachi, Almaty, Madrid and Moscow). In the figures, colour codes are used to identify 'world regions', node size varies with cities' betweenness centrality, while the edge width represents the weight of a connection whereby the 'direction' of the edge is based on the net balance of directed connections between the city nodes. To keep the figures readable, city name abbreviations are only displayed if they have an empirical betweenness of 0.005 or higher. We used the international abbreviations of the international air transport association (IATA) metropolitan codes (see appendix 1).

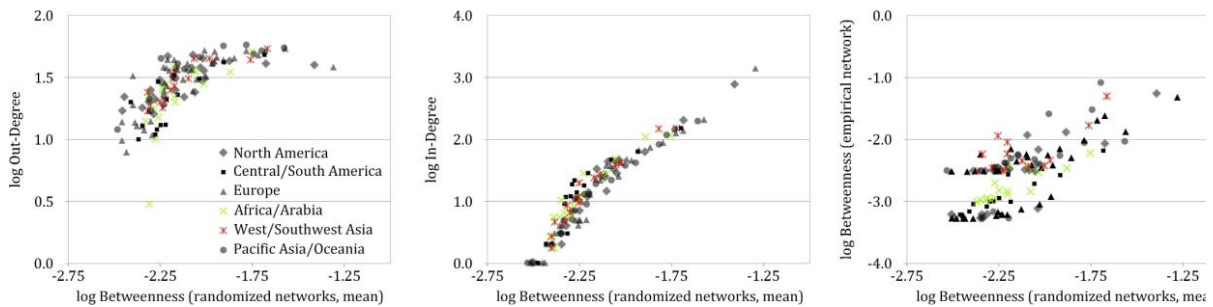

**Figure 6.** Relation between betweenness centrality in the randomized baseline networks against the empirical values derived from the linkage algorithm applied to the 2010 GaWC data

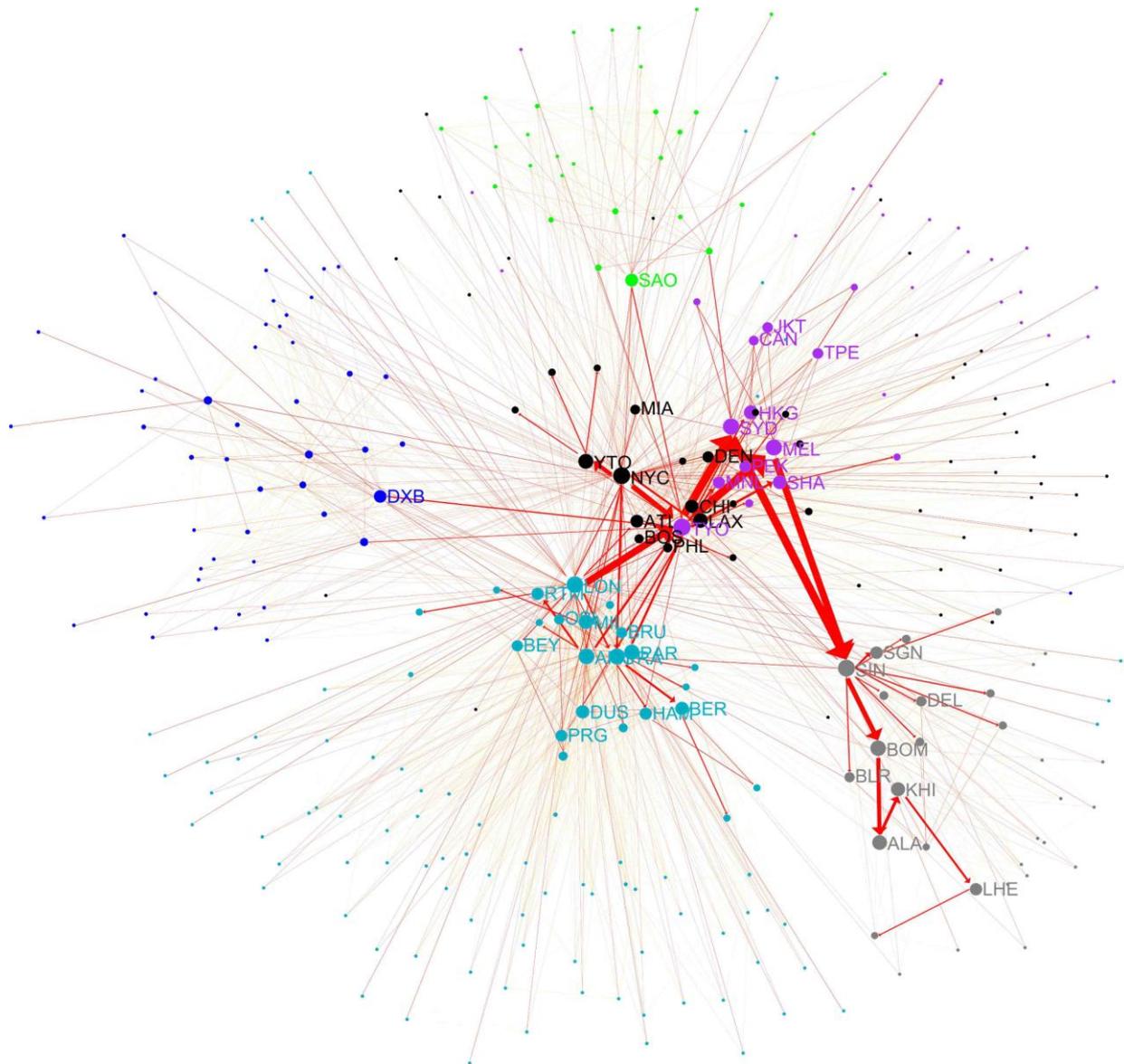

**Figure 7.** The WCN based on an application of the primary linkage method to the 2010 GaWC data (see appendix 1 for the code table)

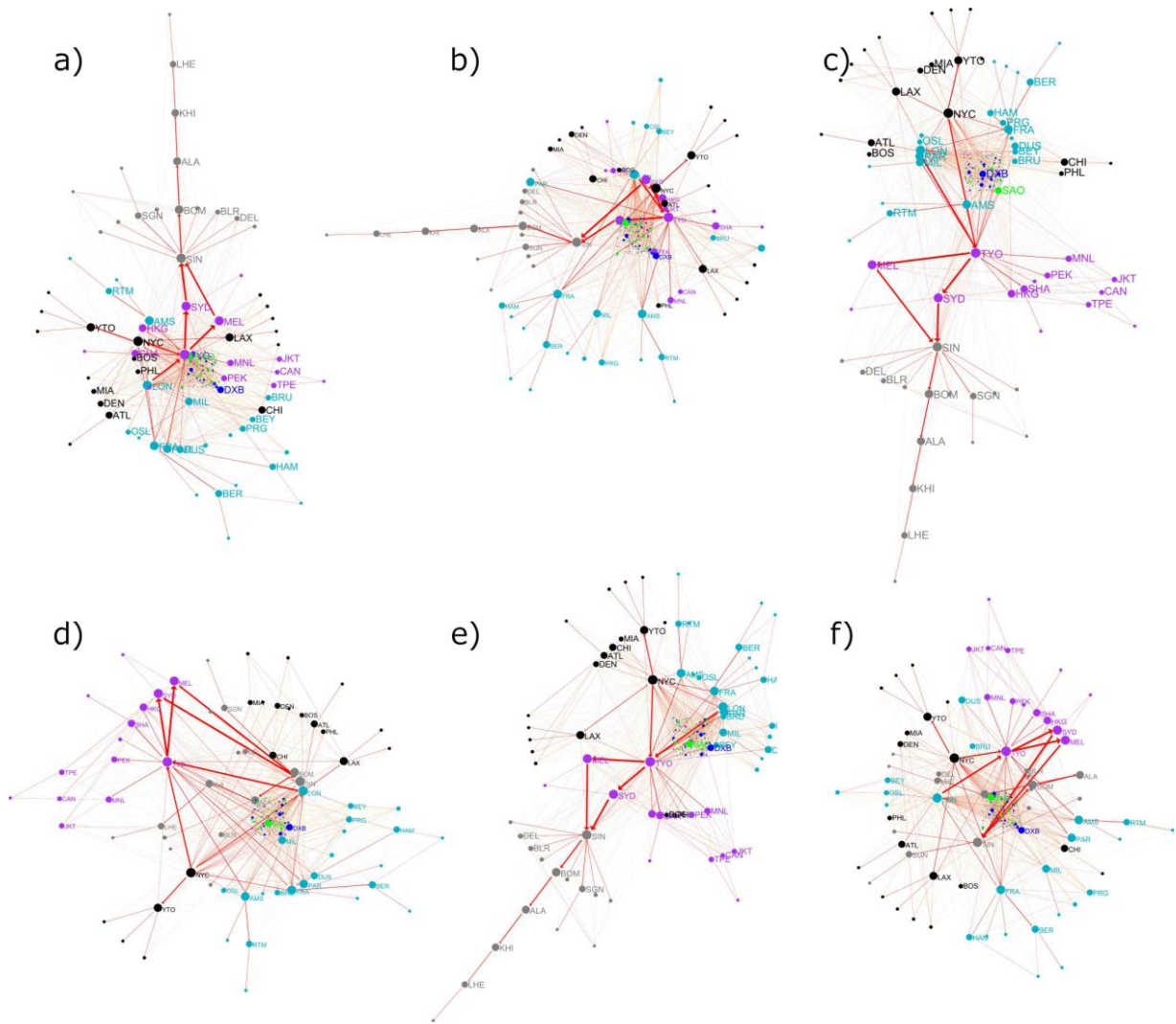

**Figure 8.** Ego network layout for a) Tokyo, b) Melbourne, c) Moscow, d) Almaty, e) Madrid, f) Karachi (see appendix 1 for the code table)

| rank (emp. Betw) | rank (rel GNC) | City | relative GNC | In-Degree | Out-Degree | Betweenness (emp) | randomized model | | emp betw / rand betw | sig. |
|---|---|---|---|---|---|---|---|---|---|---|
| | | | | | | | Betweenness (rand, mean) | Standard Deviation of the randomized mean | | |
| 1 | 6 | Tokyo | 64,8 | 146 | 52 | 0,0868 | 0,0200 | 0,0041 | 4,3 | * |
| 2 | 2 | New York | 91,7 | 833 | 40 | 0,0593 | 0,0387 | 0,0049 | 1,5 | * |
| 3 | 4 | Singapore | 69,8 | 138 | 54 | 0,0544 | 0,0215 | 0,0050 | 2,5 | * |
| 4 | 1 | London | 100,0 | 1457 | 39 | 0,0534 | 0,0491 | 0,0074 | 1,1 | |
| 5 | 7 | Sydney | 64,2 | 121 | 49 | 0,0340 | 0,0181 | 0,0038 | 1,9 | * |
| 6 | 29 | Melbourne | 51,3 | 38 | 41 | 0,0296 | 0,0109 | 0,0033 | 2,7 | * |
| 7 | 20 | Frankfurt | 57,2 | 141 | 53 | 0,0280 | 0,0210 | 0,0037 | 1,3 | * |
| 8 | 10 | Amsterdam | 61,7 | 130 | 48 | 0,0238 | 0,0192 | 0,0049 | 1,2 | |
| 9 | 18 | Mumbai | 58,2 | 136 | 44 | 0,0198 | 0,0174 | 0,0043 | 1,1 | |
| 10 | 5 | Paris | 69,5 | 217 | 55 | 0,0160 | 0,0268 | 0,0049 | 0,6 | * |
| 11 | 13 | Toronto | 60,5 | 65 | 43 | 0,0155 | 0,0134 | 0,0030 | 1,2 | |
| 12 | 19 | Los Angeles | 58,1 | 15 | 48 | 0,0141 | 0,0085 | 0,0027 | 1,7 | * |
| 13 | 125 | Almaty | 27,3 | 7 | 21 | 0,0138 | 0,0060 | 0,0018 | 2,3 | * |
| 14 | 11 | Milan | 61,5 | 90 | 52 | 0,0119 | 0,0165 | 0,0032 | 0,7 | * |
| 15 | 3 | Hongkong | 69,9 | 203 | 55 | 0,0113 | 0,0265 | 0,0052 | 0,4 | * |
| 16 | 75 | Karachi | 36,6 | 20 | 27 | 0,0111 | 0,0067 | 0,0011 | 1,7 | * |
| 17 | 9 | Shanghai | 62,2 | 84 | 58 | 0,0109 | 0,0165 | 0,0037 | 0,7 | * |
| 18 | 8 | Chicago | 63,5 | 214 | 41 | 0,0105 | 0,0211 | 0,0035 | 0,5 | * |
| 19 | 57 | Berlin | 41,5 | 13 | 37 | 0,0089 | 0,0070 | 0,0018 | 1,3 | * |
| 20 | 40 | Düsseldorf | 47,3 | 68 | 44 | 0,0089 | 0,0124 | 0,0025 | 0,7 | * |
| 22 | 15 | Sao Paulo | 59,3 | 154 | 48 | 0,0082 | 0,0207 | 0,0040 | 0,4 | * |
| 23 | 12 | Dubai | 60,9 | 115 | 51 | 0,0076 | 0,0177 | 0,0034 | 0,4 | * |
| 30 | 14 | Beijing | 59,7 | 42 | 57 | 0,0070 | 0,0124 | 0,0031 | 0,6 | * |
| 88 | 17 | Madrid | 59,0 | 40 | 52 | 0,0017 | 0,0112 | 0,0023 | 0,2 | * |
| 103 | 16 | Moscow | 59,2 | 25 | 53 | 0,0013 | 0,0098 | 0,0022 | 0,1 | * |

**Table 1.** Results of the network calculations (all cities featuring in the top 20 of either betweenness centrality based on the alternative WCN specification or GNC in the original specification are included in the table).

The results in Table 1 show that *empirical* betweenness on the one hand, and *expected* betweenness based on the null model and *GNC* on the other hand are indeed closely related (15 out of 20 cities are the same), but most certainly not the same: there are some major shifts in the ranking. Tokyo, for instance, emerges as the most central city in our respecified WCN, although it is only ranked 6[th] in Taylor's IWCN model (see Figure 7a for Tokyo's ego network). This is paired with the fact that Tokyo's empirical betweenness is also much higher than expected based on the null model, and can be attributed to the combined fact that (1) Tokyo has consistently strong connections with most other major cities such as New York, London, and Singapore, as well as (2) connecting a large group of (mainly Asian) cities to the network at large. Put differently: many of the connections passing through Tokyo are non-redundant or quasi-non-redundant for attaching these cities to the overall WCN, giving the city a much larger betweenness value than could be expected in a random network with the same degree distribution. Note that these insights regarding Tokyo's crucial position in the *overall* network cannot be drawn based on an analysis of GaWC's over-specified IWCNM.

Many of Tokyo's non-redundant or quasi-non-redundant connections pass via Singapore and Melbourne/Sydney, so that the high rankings of the latter cities can be explained along similar lines. Although Melbourne (29[th]) is deemed less connected than Sydney (7[th]) in the WCN Taylor-style, our network-wide analysis suggests that the difference between both cities is perhaps less important, as they are both playing a similar, crucial role in the WCN as a whole. That is, a large proportion of their connections is with some of the world's premier cities (especially Asia's leading cities), while both cities are literally central for keeping some other Australian/Oceanian cities connected to the WCN. Similar to Tokyo and Singapore, then, Australia's two leading cities are key nodes in the WCN.

Perhaps the most dramatic examples of this boosted betweenness in our re-specified WCN are the major connectivities for Almaty and Karachi. Although not major cities according to their GNC, both cities are important for sustaining the WCN: both cities have far fewer connections than Dubai, Madrid and Beijing, but these connections are 'strategically' important as these tend to be non-redundant for a number of cities such as Baku, Tashkent, Tbilisi, Tehran or Yerevan linking to the WCN through Almaty/Karachi and leading towards cities of high importance.

Correspondingly, some cities have less extensive levels of betweenness centrality than expected based on the baseline model and GaWC's GNC. Hong Kong, for instance, can easily be bypassed via other major cities: the city can relatively easily be circumvented via the likes of Singapore and Tokyo, making the city slightly less important than one would expect based on its cluster of globalized APS per se. Interestingly, China's two other premier cities, Beijing and Shanghai, are also statistically significant less central to the system than suggested by their GNC and the null model. An explanation here may be provided by recent qualitative research by Lai (2012), who has argued that many globalized APS firms have a three-tiered approach to China's giant space-economy. Many of these firms open offices (with slightly different functions) in each of the three cities rather than making a choice between them. Although this makes China's three premier cities well connected in the office networks of APS firms, these connections are, in relative terms, redundant in the sense that one could possibly call up an office in one of the other two cities. Put differently: each of the three Chinese cities can on average be more easily bypassed than, say, Singapore, and the latter city thus boasts a more crucial position in the overall network structure.

Similarly, the redundancy of some of the connections of Moscow and Madrid is hindering both European cities from being more prominently ranked: the fact that there are many other connections leading to the same target city node implies that both cities can be bypassed relatively easily via other European cities. In the IWCNM, the extensive clusters of APS in Madrid and Moscow lead to a major GNC as everything-is-connected-to-everything, but here we suggest that this IWCNM approach possibly overvalues these cities' importance, as most of their connections are unneeded for reproducing the spatial structure of the WCN; indeed, both cities are offering only weak regional hub functions for less connected cities in their region.

A different perspective on the top 20 global cities is shown in figure 9. It is summing up the information in table 1 and maps the empirical betweenness against the randomized mean betweenness. The error bars are showing the standard deviation SD of the betweenness estimates. All cities that are below the dashed 45-degree line are showing higher empirical betweenness values than expected by chance. For those cities above the line, the values are lower than expected. However, if the vertical error bars cross the line, the differences are not sufficiently large enough, i.e. significant (such as in the case of London).

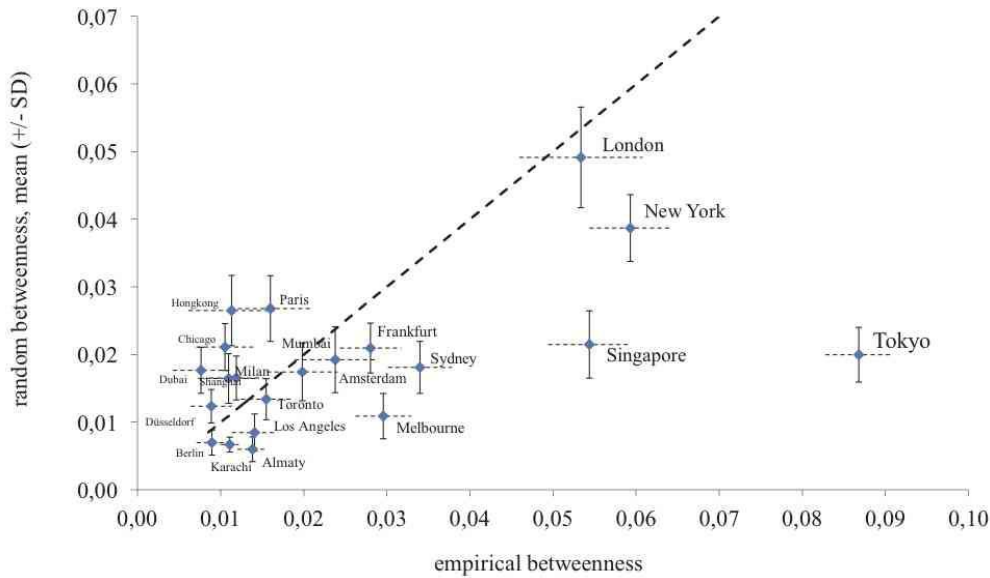

**Figure 9**. Empirical betweenness vs. random estimates for the Top 20 cities

The dashed horizontal error bars show the random mean SD mapped to the empirical values to indicate for significant ranking position, i.e. if the lines of any two cities are overlapping one another, the cities' ranking position is not significantly different, such as in the cases of London, New York and Singapore. However, Singapore is the most surprising one in this group of three, because of the large deviation between empirical and randomized betweenness values. Moreover, it tells that the ranking position of Tokyo is pretty solid, given the large margin to the followers. Likewise, Sydney is significantly better positioned than Amsterdam, whereas Amsterdam's rank and Melbourne's rank are not significantly different from one another.

This focus on 'major' connectivities clearly shows the potential of our approach, and we can therefore – to conclude – turn to what are perhaps less intuitive but nonetheless equally interesting findings of a set of second-tier cities. Table 2 lists the empirical betweenness centrality, the betweenness centrality emanating from the null model, and GNC for second-ranked European cities. The table is split in two halves: a group of cities that loses relevance when compared to GaWC's GNC ("losing cities" such as Madrid and Warsaw), and a group of cities that have a significantly higher betweenness connectivity than expected given their relative GNC ("gaining cities" such as Oslo and The Hague). In general, the gaining cities have the tendency of having high numbers of incoming connections relative to their outgoing connections. This implies that these cities assemble relatively high office functions in firm networks that are quite important for the overall network functioning. Although they do not necessarily possess the *highest-level* functions in the office networks, these cities are mediating flows between hierarchical levels and therefore for cities that are *otherwise less accessible* in the WCN. Moreover, these gaining cities are more often than not hosting the most important offices of APS firms in their country.

The obverse interpretation holds for the losing cities, which are comparatively well-connected cities that are nonetheless not offering mediating functions between cities in the WCN. In line with Moscow/Madrid examples, then it can be said that their overall connectivity is less exclusive, i.e. they are connected to a large variety of cities outside their own region without offering non-redundant paths.

| #rank (emp. Betw) | #rank (rel GNC) | City | relative GNC | In-Degree | Out-Degree | Betweenness (emp) | randomized model ||| sig. |
| | | | | | | | Betweenness (rand, mean) | Standard Deviation of the randomized mean | emp betw / rand betw | |
|---|---|---|---|---|---|---|---|---|---|---|
| 38 | 63 | Oslo | 39,80 | 23 | 31 | 0,00579 | 0,0075 | 0,0019 | 0,77 | |
| 43 | 36 | Munich | 48,90 | 46 | 46 | 0,00504 | 0,0114 | 0,0030 | 0,44 | * |
| 44 | 56 | Copenhagen | 41,80 | 38 | 31 | 0,00501 | 0,0089 | 0,0016 | 0,56 | * |
| 51 | 43 | Dublin | 46,50 | 43 | 33 | 0,00463 | 0,0100 | 0,0027 | 0,46 | * |
| 63 | 86 | Stuttgart | 34,00 | 12 | 27 | 0,00411 | 0,0061 | 0,0012 | 0,67 | * |
| 67 | 94 | Cologne | 32,10 | 8 | 25 | 0,00409 | 0,0059 | 0,0013 | 0,69 | * |
| 70 | 167 | Bologna | 21,70 | 4 | 14 | 0,00408 | 0,0050 | 0,0009 | 0,82 | * |
| 72 | 194 | Hannover | 18,80 | 5 | 11 | 0,00406 | 0,0051 | 0,0008 | 0,80 | * |
| 73 | 169 | Leipzig | 21,70 | 2 | 13 | 0,00406 | 0,0042 | 0,0012 | 0,96 | |
| 76 | 180 | The Hague | 20,50 | 3 | 12 | 0,00406 | 0,0046 | 0,0013 | 0,89 | |
| 88 | 17 | Madrid | 59,00 | 40 | 52 | 0,00174 | 0,0112 | 0,0023 | 0,16 | * |
| 103 | 16 | Moscow | 59,20 | 25 | 53 | 0,00129 | 0,0098 | 0,0022 | 0,13 | * |
| 106 | 34 | Zurich | 49,80 | 22 | 41 | 0,00115 | 0,0079 | 0,0016 | 0,15 | * |
| 109 | 85 | Geneva | 34,50 | 7 | 29 | 0,00103 | 0,0060 | 0,0013 | 0,17 | * |
| 156 | 41 | Barcelona | 47,30 | 5 | 39 | 0,00092 | 0,0059 | 0,0021 | 0,16 | * |
| 157 | 53 | Stockholm | 42,30 | 10 | 36 | 0,00091 | 0,0063 | 0,0018 | 0,14 | * |
| 158 | 27 | Vienna | 52,00 | 5 | 44 | 0,00091 | 0,0067 | 0,0019 | 0,14 | * |
| 159 | 49 | Tel Aviv | 43,50 | 1 | 33 | 0,00089 | 0,0040 | 0,0020 | 0,22 | * |
| 161 | 45 | Istanbul | 44,90 | 5 | 38 | 0,00088 | 0,0061 | 0,0018 | 0,14 | * |
| 163 | 37 | Rome | 48,60 | 4 | 45 | 0,00086 | 0,0061 | 0,0021 | 0,14 | * |
| 164 | 33 | Warsaw | 49,90 | 5 | 44 | 0,00086 | 0,0062 | 0,0016 | 0,14 | * |

**Table 2.** Building blocks of differences in ranking of medium-connected European cities

## 6 Concluding remarks

In this paper, we have addressed some of the limitations of Taylor's (2001) IWCNM specification with the purpose of devising an actual network analysis of the WCN. We have specified an alternative one-mode projection function, which results in sparser adjacency matrix that nonetheless retains the basic contours of the IWCNM-produced urban network. The resulting inter-city matrix can – in contrast to the WCN produced by the unaltered IWCNM – be examined with network analysis techniques. Here we have analysed cities' positions in the network at large by comparing their empirical betweenness centrality with a randomized null-model, but other network analysis options are of course possible. We discussed the empirical advantages of our approach through a number of notable examples, i.e. cities that rank much higher/lower on betweenness centrality than on IWCNM-produced centrality measures. The added insight includes, for instance, the identification of cities that play a major role in APS firms' office networks because their links to major cities are non-redundant (e.g., Almaty), while other cities are less important because their position in APS firms' office networks are redundant from the perspective of the network at large (e.g. Hong Kong, which can be by-passed through Singapore or Tokyo).

Nonetheless, it is important to stress that the relevance of our analysis primarily lies in showing how Taylor's (2001) approach can be extended to arrive at actual network analyses of the WCN. That is, it is clear that our framework also impacts the results so that some of the differences may be technical artefacts of the regionalization etc. rather than 'actual' results. Although a city's empirical degree centrality and GNC are nearly perfectly correlated, the linkage procedure may nonetheless have an impact. For instance, it is possible that adding Pacific Asia and Oceania in a single world region may explain Melbourne's and Sydney's high betweenness centrality, and an obvious avenue for further research is therefore assessing the impact of this (or other) particular one-mode projections. This type of sensitivity analysis is subject to further research and needs to be extended to the sensitivity of the ranking results, as they have proven not always to be significantly different for the midfield of the global city ranking (cf. Figure 9). Also from a methodological point of view, our approach and data needs to be evaluated against such systems as global transportation/airline networks or global science networks.

From a theoretical/conceptual perspective, our results suggest to further shape the idea of geographical holes in the world city network, i.e. cities that are obviously important for the systemic structure without being overwhelmingly well connected. This avenue of research would link to the discussion in Liefner and Hennemann (2011), who combine the network perspective and the future developmental perspective of agglomerations and attribute territorial lock-ins and economic decline to missing structural holes capacities of an agglomerations' network position. The panel-like structure of the GaWC research network data on advanced producer service firms would allow for empirical testing of this notion. For the 2010 data that was used here, we found Madrid and Moscow in such potentially disadvantageous positions. Both cities are rich in connectivity, while offering only weak regional hub function for less connected cities in their neighbourhood (both, regional and network related). This conception would introduce an inherent dynamic component into the idea of WCN, pushing it towards current theoretical discussions in human geography such as evolutionary economic geography in a global perspective.

**References**

Andriani P, McKelvey B, 2009, "From Gaussian to Paretian thinking: causes and implications of power laws in organizations" *Organization Science* **20** (6) 1053-1071

Barabási A-L, Albert R, 1999, "Emergence of scaling in random networks" *Science* **286** (5439) 509-512

Derudder B, Taylor P J, Witlox F, Catalano G, 2003, "Hierarchical tendencies and regional patterns in the world city network: a global urban analysis of 234 cities" *Regional Studies* **37** (9) 875-886

Derudder B, Taylor P J, 2005, "The cliquishness of world cities" *Global Networks* **5** (1) 71-91

Gilsing V, Nooteboom B, Vanhaverbeke W, Duysters G, van den Oord A, 2008, "Network embeddedness and the exploration of novel technologies: technological distance, betweenness centrality and density" *Research Policy* **37** (10) 1717-1731

Godfrey B J, Zhou Y, 1999, "Ranking world cities: multinational corporations and the global urban hierarchy" *Urban Studies* **20** (3) 268-281

Hennemann S, Rybski D, Liefner I, 2012, "The myth of global science collaboration - Collaboration patterns in epistemic communities" *Journal of Informetrics* **6** (2) 217-225

Hennemann S, 2012, "Evaluating the performance of geographical locations within scientific networks using an aggregation – randomization – re-sampling approach (ARR)" *Journal of the American Society for Information Science and Technology* **63** (forthcoming)

Hoyler M, Watson A, 2012, "Global media cities in transnational media networks" *Tijdschrift voor Economische en Sociale Geografie* (forthcoming)

Keeling D J, 1995, "Transportation and the world city paradigm", in *World Cities in a World-System* Eds P L Knox, P J Taylor (Cambridge University Press, Cambridge) pp 115-131

Latapy M, Magnien C, Del Veccio N, 2008, "Basic notions for the analysis of large affiliation networks / bipartite graphs" *Social Networks* **30** (1) 31-49

Liefner I, Hennemann S, 2011, "Structural holes and new dimensions of distance: The spatial configuration of the scientific knowledge network of China's optical technology sector" *Environment and Planning A* **43** (4) 810-829

Lai K, 2012, "Differentiated markets: Shanghai, Beijing and Hong Kong in China's financial centre network" *Urban Studies* **49** (6) 1275-1296

Liu X, Derudder B, 2012, "Two-mode networks and the interlocking world city network model: a reply to Neal" *Geographical Analysis* **44** (2) 171-173

Liu X, Derudder B, Liu Y, 2012, "Modelling the intercity corporate network: an exponential random graph modelling approach" (Under review)

Liu X, Taylor P J, 2011, "A robustness assessment of GaWC global network connectivity ranking" *Urban Geography* **32** (8) 1227-37

Maslov S, Sneppen K, 2002, "Specificity and stability in topology of protein networks" *Science* **296** (5569) 910-913

Meijers E, Burger M, van Oort F, 2012, "Identifying networks in the polycentric mega-city region: a review of current approaches and suggestions for an alternative approach based on complementarity and related variety", paper presented at AAG2012, 24-28 February, http://meridian.aag.org/callforpapers/program/AbstractDetail.cfm?AbstractID=42096

Neal Z, 2008, "The duality of world cities and firms: networks, hierarchies, and inequalities in the global economy" *Global Networks* **8** (1) 94-115

Neal Z, 2012a, "Structural determinism in the interlocking world city network" *Geographical Analysis* **44** (2) 162-170

Neal Z, 2012b, "Intercity linkages in the interlocking world city network: how strong is strong enough?", paper presented at AAG2012, 24-28 February, http://meridian.aag.org/callforpapers/program/AbstractDetail.cfm?AbstractID=41241

Nordlund C, 2004, "A critical comment on the Taylor approach for measuring world city interlocking linkages" *Geographical Analysis* **36** (3) 290-296

Robinson J, 2002, "Global and world cities: a view from off the map" *International Journal of Urban and Regional Research* **26** (3) 531-554

Sassen S, 1991 *The Global City: New York, London, Tokyo* (Princeton University Press, Princeton)

Smith R G, 2012, "Beyond the global city concept", Globalization and World City research network Research Bulletin 390, http://www.lboro.ac.uk/gawc/rb/rb390.html

Taylor P J, 2001, "Specification of the world city network" *Geographical Analysis* **33** (2) 181-194

Taylor P J, 2004, "Reply to 'a critical comment on the Taylor approach for measuring world city interlock linkages'" *Geographical Analysis* **36** (3) 297-298

Taylor P J, 2005, "New political geographies: global civil society and global governance through world city networks" *Political Geography* **24** (6) 703-730

Taylor P J, Catalano G, Walker D R F, 2002, "Measurement of the world city network" *Urban Studies* **39** (13) 2367-2376

Taylor P J, Ni P, Derudder B, Hoyler M, Huang J, Witlox F, 2010 *Global Urban Analysis: A Survey of Cities in Globalization* (Routledge, London)

Taylor P J, Derudder B, Hoyler M, Ni P, 2012, "New regional geographies of the world as practised by leading advanced producer service firms in 2010" *Transactions of the Institute of British Geographers* (accepted for publication)

van Wijk B C M, Stam C J, Daffertshofer A, 2010, "Comparing brain networks of different size and connectivity density using graph theory" *Public Library of Science ONE* **5** (10) e13701

Watts D J, Strogatz S H, 1998, "Collective dynamics of 'small-world' networks" *Nature* **393** (6684) 440-442

Appendix 1: List of IATA-City/Metropolitan codes that have been used in figures 7 and 8

| Code | City | Code | City |
|------|------|------|------|
| ALA | Almaty | LHE | Lahore |
| AMS | Amsterdam | LON | London |
| ATL | Atlanta | MEL | Melbourne |
| BER | Berlin | MIA | Miami |
| BEY | Beirut | MIL | Milan |
| BLR | Bangalore | MNL | Manila |
| BOM | Mumbai | MUC | Munich |
| BOS | Boston | NYC | New York |
| BRU | Brussels | OSL | Oslo |
| CAN | Guangzhou | PAR | Paris |
| CHI | Chicago | PEK | Beijing |
| CPH | Copenhagen | PHL | Philadelphia |
| DEL | New Delhi | PRG | Prague |
| DEN | Denver | RTM | Rotterdam |
| DUS | Düsseldorf | SAO | Sao Paulo |
| DXB | Dubai | SGN | Ho Chi Minh City |
| FRA | Frankfurt | SHA | Shanghai |
| HAM | Hamburg | SIN | Singapore |
| HKG | Hongkong | SYD | Sydney |
| JKT | Jakarta | TPE | Taipei |
| KHI | Karachi | TYO | Tokyo |
| LAX | Los Angeles | YTO | Toronto |